\documentclass[doublecol]{epl2}

\usepackage{amsmath}
\usepackage{amssymb}
\usepackage{graphicx}
\usepackage{dcolumn}
\usepackage{bm}
\usepackage{color}

\title{Luttinger parameters and momentum distribution function
for the half-filled spinless fermion Holstein model: A DMRG approach}
\shorttitle{TLL parameter and momentum distribution function for the
Holstein model}
\author{S.\ Ejima \and H.\ Fehske}
\institute{Institut f{\"ur} Physik,
          Ernst-Moritz-Arndt-Universit{\"a}t Greifswald,
          17489 Greifswald,
          Germany}
\date{\today}
\abstract{We reexamine the nature of the metallic phase of the
one-dimensional half-filled Holstein model of spinless fermions.
To this end we determine the Tomonaga-Luttinger-liquid 
correlation parameter  
$K_\rho$ 
by large-scale density-matrix renormalisation-group 
(DMRG) calculations, exploiting (i) the leading-order scaling relations
between the ground-state energy and the single-particle excitation gap
and (ii) the static charge structure factor in the long-wavelength
limit. While both approaches give almost identical results for 
intermediate-to-large phonon frequencies, we find contrasting 
behaviour in the adiabatic regime: (i) $K_\rho>1$ (attractive) 
versus (ii) $K_\rho<1$ (repulsive).
The latter result for the correlation exponent is corroborated 
by data obtained for the momentum distribution function $n(k)$, which puts the 
existence of an attractive metallic state in the spinless
fermion Holstein model into question. We conclude that the scaling 
relation must be modified in the presence of electron-phonon 
interactions with noticeable retardation.}

\begin{document} 
\maketitle

%
During the last three decades we have seen a constant growth
of experimental realizations of one-dimensional (1D) materials.
Nowadays the progress in nanotechnology allows to manufacture isolated
carbon nanotubes or quantum wires~\cite{PN93}. But there are also bulk
materials with quasi-1D structures inside.
Famous examples are conjugated polymers, charge transfer salts, 
halogen-bridged transition metal complexes, ferroelectric perovskites, 
spin Peierls compunds, molecular metals or organic 
superconductors~\cite{IYS73}. 
The apparent diversity of physical properties observed for different 
material classes has its seeds in the strong competition between   
the itinerancy of the electronic charge carriers on the one hand and the 
electron-electron and electron-lattice interactions on the other hand.  
The latter tend to establish insulating spin-density-wave or 
charge-density-wave (CDW) ground states, respectively,
at least for commensurate band 
fillings~\cite{Mo90,Pe55,TC03,FWHWB04,CH05,TAA05}.
Interactions have drastic effects in 1D systems
compared to higher dimensions. Most notably one observes 
a ``collectivisation'' of any excitation.  As a consequence, 
for fermionic systems, the usual Fermi liquid description breaks 
down~\cite{Gi07}. Luttinger liquid  theory provides an adequate 
compensation~\cite{Lu63}. It tells us that all ground-state, spectral 
and thermodynamic properties of a Luttinger liquid  are basically 
controlled by a few (non-universal) parameters. This result
can be used in the following way. Starting from a specific 
microscopic model, one can try to compute certain (thermodynamic) 
quantities exactly, e.g. for finite systems by elaborate numerical 
techniques, and afterwards extract the Luttinger liquid parameters, e.g.  
the charge correlation exponent $K_\rho$ and charge velocity $u_\rho$, 
out of them. 
Advantageously these parameters, describing 
the overall low-energy physics of our system, are much less sensitive to 
finite-size effects than the correlation functions themselves.  
Of course, the concept of a Luttinger liquid has to be taken
as a starting point to study more complex situations, comprising e.g.
the lattice degrees of freedom or disorder effects. Then the Luttinger 
parameters become effective parameters, which characterise very basic 
properties of the system, such as an attractive  ($K_\rho>1$)  or repulsive
 ($K_\rho>1$) interaction between the particles~\cite{Gi07,Vo95}. 

In this respect, focusing on the coupling of charge carriers to the vibrations 
of a deformable lattice, the so-called Holstein model 
of spinless fermions (HMSF)~\cite{Ho59a},  
\begin{eqnarray}
 {\cal H}
  &=& -t\sum_{j}
       \left(
	c^\dagger_{j} c^{\phantom{\dagger}}_{j+1} +
        {\rm h.c.}
       \right)
      +\omega_0\sum_j b^{\dagger}_j b^{\phantom{\dagger}}_{j}
                        \nonumber \\
  & & - g\omega_0\sum_{j}
   \left(  
    b^{\dagger}_j + b^{\phantom{\dagger}}_{j}
   \right) \left(n_j-\frac{1}{2}\right)\,,
\label{eqn:H-Holstein}
\end{eqnarray}
 is particularly rewarding to study. 
It accounts for a tight-binding electron band ($\propto t$), a 
local electron-phonon (EP) interaction ($\propto g$), and the 
energy of the phonon subsystem in harmonic approximation.  
In eq.~(\ref{eqn:H-Holstein}), $c^{\dagger}_{j}$ ($c^{}_{j}$) 
creates (annihilates) an electron at Wannier site $j$ of a 1D lattice 
with $N$ sites, $b^{\dagger}_{j}$ ($b_{j}$) are the corresponding 
bosonic operators for a dispersionless optical phonon, and $\omega_0$
is the frequency of such an internal (e.g. molecular) 
vibration.   

Despite its seemingly simplicity, the 1D HMSF is not exactly solvable.
It is generally accepted, however, that the model exhibits a 
quantum phase transition from a metal to a CDW insulator
at half-filling, when the EP coupling $g$ increases at 
fixed $\omega_0>0$.\footnote{We consider the half-filled band case 
hereafter, i.e., $\tfrac{1}{N}\sum_{j}\langle 
	c^\dagger_{j} c^{\phantom{\dagger}}_{j}\rangle=\tfrac{1}{2}$, 
and take $t=1$ as energy unit.}
The CDW phase above $g_{\rm c}(\omega_0)$ is connected to a (Peierls) distortion 
of the lattice, and can be classified 
as traditional band insulator and polaronic superlattice, respectively, 
in the adiabatic ($\omega_0\ll 1$) and anti-adiabatic 
($\omega_0 \gg 1$) regimes~\cite{FHW00,HWBAF06}. 
A wide range of analytical and numerical methods have been applied to map
out the phase diagram of the HMSF in the whole $g-\omega_0$ 
plane~\cite{HF82,ZFA89,MHM96,BMH98,FHW00,SHBWF05}, 
with significant differences in the region of small-to-intermediate
phonon frequencies. The results agree in the anti-adiabatic strong EP 
coupling limit ($\omega_0 \to \infty$, $g>1$), where the 
HMSF possesses XXZ-model physics. There a Kosterlitz-Thouless type 
transition~\cite{KT73} occurs at the spin isotropy point,
with $K_\rho$ reaching 1/2 from above   
at the transition point~\cite{HF82,BMH98}. 

In the first instance, however, the correlation exponent $K_\rho$ 
can be used to characterise the metallic phase itself. 
According to Haldane's conjecture~\cite{Ha80}, 
a 1D gapless (metallic) system of interacting fermions should belong
to the Tomonaga-Luttinger liquid (TLL) universality class~\cite{To50,Lu63}. 
For a TLL of spinless fermions, the ground-state energy $E(N)$ 
and the one-particle charge excitation gap $\Delta_{c_1}$
of a finite system with $N$ sites scale to leading order as~\cite{Vo95,Ca84}:
\begin{eqnarray}
&& \frac{E(N)}{N} = \varepsilon(\infty)-\frac{\pi}{3}
    \frac{u_\rho}{2}\frac{1}{N^2},
\label{eqn:krho1}
  \\
&& \Delta_{c_1}=E^{\pm}(N)-E(N) 
     =  \pi\frac{u_\rho}{2}\frac{1}{K_\rho}\frac{1}{N}.
\label{eqn:urho1}
\end{eqnarray}
Here, $\varepsilon(\infty)$ denotes the energy density of the infinite
system with $N/2$ electrons, $E^{\pm}(N)$ are the ground-state energies
with $\pm 1$ fermion away from half filling, and $u_\rho$ is the 
renormalised charge velocity.\footnote{The TLL scaling 
relations (\ref{eqn:krho1}) and (\ref{eqn:urho1}) were also derived for 
spinful systems~\cite{FK90}, and e.g. used in order to compute the central
charge in the framework of the $t$--$J$ model~\cite{Og91}.} Using 
these equations from field theory,  in the past, $K_\rho$ and $u_\rho$ were 
determined for the HMSF on finite clusters by various exact numerical  
techniques~\cite{MHM96,BMH98,FHW00,WFWB00}.
Interestingly the TLL phase seems to split into two different 
regions: For large phonon frequencies the effective fermion-fermion 
interaction is repulsive ($K_\rho<1$), while it is attractive 
($K_\rho>1$) for small frequencies~\cite{FHW00,WFWB00}. 
In the former (anti-adiabatic) 
regime the kinetic energy ($\propto u_{\rho}$) is strongly reduced 
and the charge carriers behave like (small) polarons~\cite{Fi75,FT07}. 
By contrast  the mass renormalisation is rather weak in the adiabatic 
regime~\cite{WFWB00}. The size of the phonon frequency 
also significantly affects the electron and phonon spectral 
functions~\cite{SHBWF05,HWBAF06,SHB06}. 
 
\begin{figure}[t]
\includegraphics[width=1.0\linewidth]{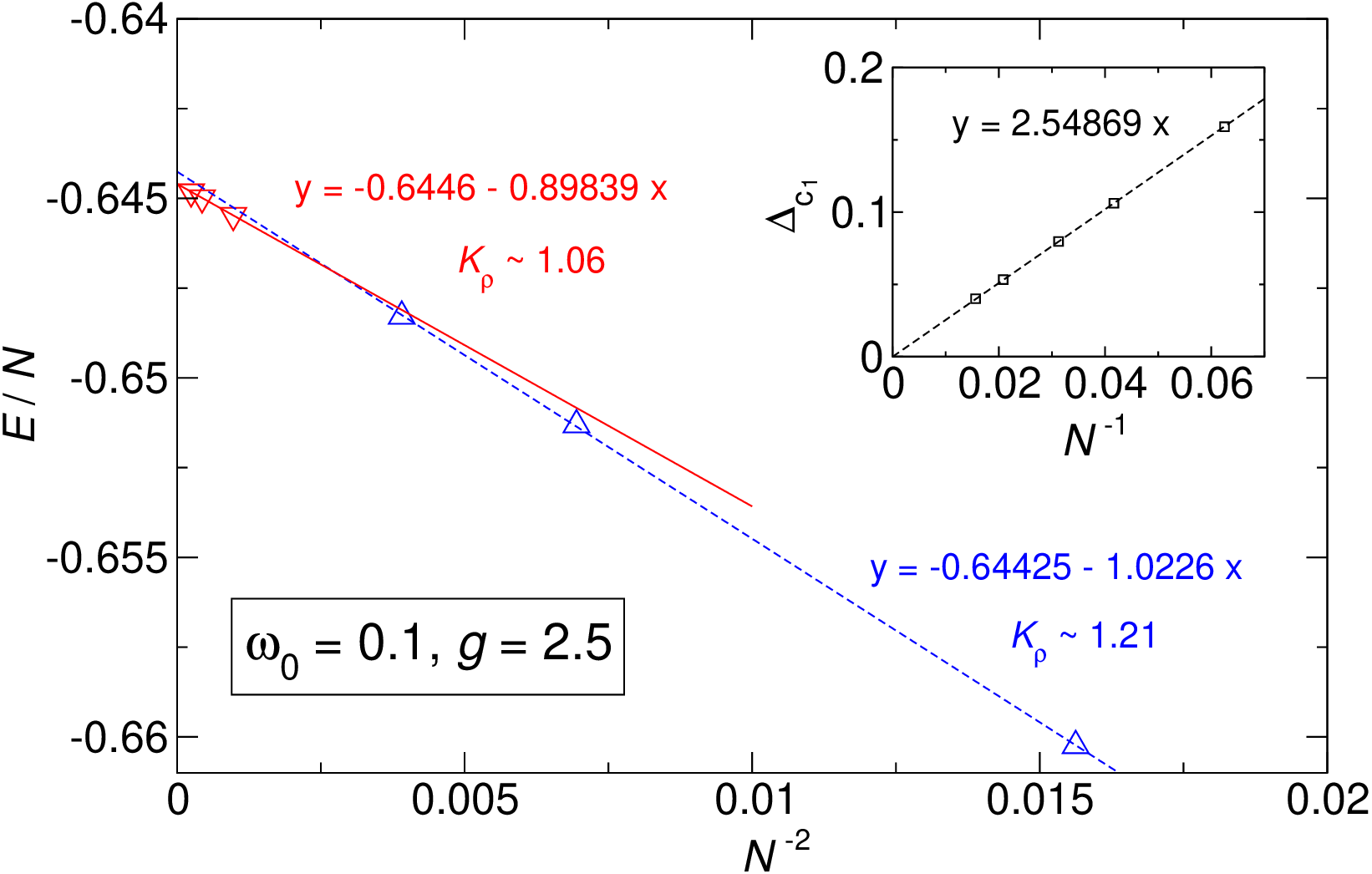}
\caption{(Colour online) Finite-size scaling of the ground-state energy
 $E(N)$ and the one-particle charge excitation gap $\Delta_{c_1}$ (inset) 
 in the spinless Holstein model at half filling. Results are obtained 
 by DMRG for $\omega_0=0.1$ and $g=2.5$.
 The linear equations give the coefficients of a straight-line fit  
 to the scaling relations (\ref{eqn:krho1}) and (\ref{eqn:urho1}).
}
\label{fig:krho-Holstein2}
\end{figure}

The existence of an attractive TLL ($K_\rho >1$) in the HMSF is by no means 
obvious however. Although retardation effects might lead to an effective 
attraction between electrons at small $\omega_0/t$ (i.e., a second 
electron may take the advantage of the lattice distortion left by the 
first one),  it has been pointed out that such an interaction is ineffective 
in the case of spinless fermions for small EP 
couplings because of the Pauli exclusion principle~\cite{HF82}.\footnote{Note 
this argument does not hold for the spinful Holstein model.}
Furthermore, if $K_\rho$ would increase with increasing EP coupling
at small $\omega_0$, as indicated by different numerical 
studies~\cite{FHW00,WFWB00} exploiting 
equations~(\ref{eqn:krho1}) and (\ref{eqn:urho1}), 
how could we detect the phase transition 
from $K_\rho \to 1/2$ in the adiabatic regime?
Of course, equations~(\ref{eqn:krho1}) and (\ref{eqn:urho1})
are leading-order expressions only, and nonlinear correction 
terms have to been taken into account in order to obtain accurate data 
for $g_c$~\cite{BMH98}. This particularly applies to the adiabatic region.
According to table~III in Ref.~\cite{BMH98} the difference between
the $g_c$ determined with and without nonlinear correction terms 
adds up to more than 3\% for $\omega_0=0.1$,  whereas it is only 0.4\% 
for $\omega_0=10$.
Actually the charge velocity $u_\rho$ depends strongly on the system's size
as shown in figure~\ref{fig:krho-Holstein2} obtained by a 
density matrix renormalisation group (DMRG) calculation.
Extrapolating the ground-state energies $E(N)$ for $N=8$, 12 and 16,  
the charge velocity $u_\rho$ can be estimated as $u_\rho/2\sim0.977$,
so that $K_\rho\sim 1.21$ from the finite-size scaling of $\Delta_{c_1}$, 
while taking the ground-state energies for $N=32$, 48 and 64, the extracted 
value of $K_\rho$ reduces to 1.06 ($u_\rho/2 \sim 0.858$).
Moreover we are faced with the difficulty that the single particle excitation 
gap seems to scale to zero (see inset of 
figure~\ref{fig:krho-Holstein2} for $\omega_0=0.1$, $g=2.5$), 
i.e. $\Delta_{c1}$ gives no signal for a pairing instability. 
Because of this situation it is highly desirable to find a reliable and 
numerical efficient method for calculating the correlation exponent $K_\rho$ 
with high precision in the whole TLL regime. 

Recently Ejima {\it et al.}~\cite{EGN05} have shown that $K_\rho$ 
can be determined for fermionic models accurately in an 
alternative way:  By a DMRG calculation of the charge structure factor
for systems with open boundary conditions. 
The approach was extended to coupled fermion-boson systems 
and has been used, e.g., to analyse  
the metal-insulator transition points
in a model with boson affected transport, for both small and large
boson frequencies~\cite{EHF09}. 

In this work we adapt this calculation scheme to the 
1D Holstein model of spinless fermions~(\ref{eqn:H-Holstein}) and 
reexamine, in particular, the possible existence of 
a metallic phase with attractive interaction. To this end 
we compute, in a first step, the static charge structure factor
\begin{eqnarray}
 S_c(q)=\frac{1}{N}\sum_{j,l} e^{iq(j-l)}
         \left\langle 
          \left(
	   c_j^\dagger c_j^{{\phantom\dagger}}-\frac{1}{2}
          \right)
          \left(
	   c_l^\dagger c_l^{{\phantom\dagger}}-\frac{1}{2}
          \right)
         \right\rangle\,,
\end{eqnarray}
and extract, in a second step, the TLL correlation exponent $K_\rho$, being 
proportional to the slope of $S_c(q)$ in the long-wavelength limit~\cite{CSC99,EGN05,CH05}:
\begin{eqnarray}
 K_\rho=\pi\lim_{q\to 0^+}\frac{S_c(q)}{q}, \ \ \ q=\frac{2\pi}{N},\ \ \
  N\to \infty\,.
\label{eqn:Krho-Sc}
\end{eqnarray}
Moreover we calculate the momentum distribution function 
for the HMSF and, having accurate data for $K_\rho$ at hand, 
analyse the results within a TLL description, also in relation 
to the corresponding results for the half-filled spinless $t$-$V$ model. 
When treating coupled 1D fermion-boson systems by DMRG we employ the 
pseudo-site approach \cite{JW98b} 
which maps a bosonic site, 
containing $2^{n_b}$ states, exactly to $n_b$ pseudo-sites.
For the numerics presented below we have taken into account up to $n_b=5$
pseudo-sites, so that the $n_b$-th local boson density is always
smaller than $10^{-8}$. In addition we kept $m=1200$ density-matrix
eigenstates. Then the discarded weight was always smaller 
than $1.0\times10^{-10}$.

\begin{figure}[tb]
\includegraphics[width=1.0\linewidth]{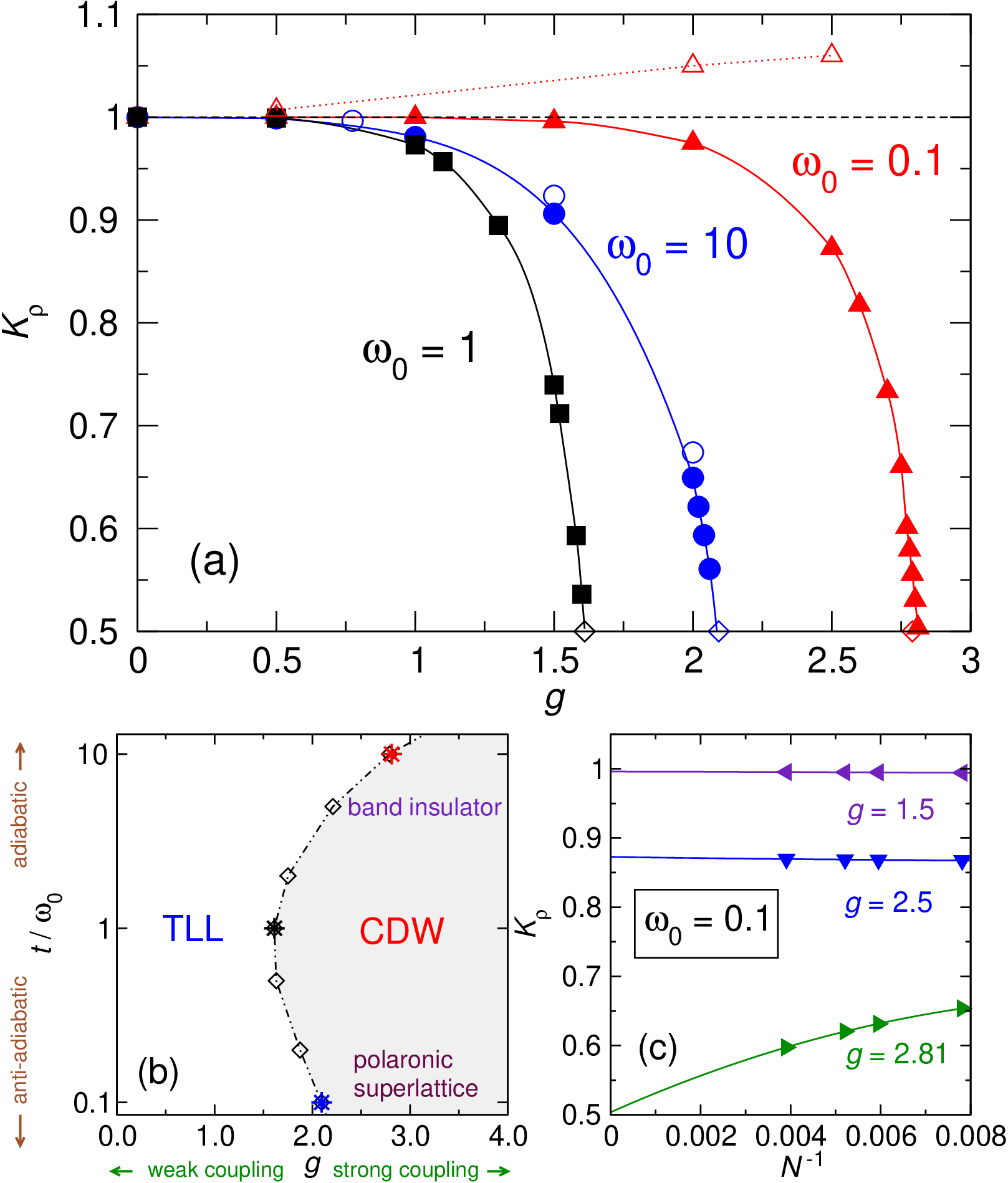}
\caption{(Colour online) Panel (a): TLL parameter $K_\rho$ in the spinless
 Holstein model at half filling. Closed symbols are obtained 
 via $S_c(q)$ from eq.~(\ref{eqn:Krho-Sc}) for $\omega_0=0.1$ (triangles), 
 $1$ (squares), and 10 (circles). $K_\rho$ obtained from the 
 scaling relations~(\ref{eqn:krho1})
 and (\ref{eqn:urho1}) are included as open symbols.
 Lines are guides to the eye.
 Panel (b): Ground-state phase diagram of the 1D half-filled spinless 
 Holstein model according to Refs.~\cite{BMH98,FHW00,HWBAF06}.
 Stars denote the phase transition points obtained from $K_\rho = 1/2$ 
 in (a).  
 Panel (c): $K_\rho$ as a function of the inverse system size 
 at various EP couplings $g=1.5$, 2.5, and 2.81 
 for $\omega_0=0.1$ (adiabatic regime). 
 Lines are polynomial fits.
}
\label{fig:Krho-PD}
\end{figure}

Figure~\ref{fig:Krho-PD} (a) presents $K_\rho$ obtained from
eq.~(\ref{eqn:Krho-Sc}) for various phonon frequencies. Note that
data points in (a) represent $K_\rho$-values extrapolated
to the infinite system at fixed $(g,\omega_0)$ [cf., e.g., panel (c)].   
For intermediate-to-large phonon frequencies we find $K_\rho(g)<1$
for all $g$, but an appreciable reduction of $K_\rho$ takes place above $g=1$ 
only. The strong decrease of $K_\rho$ and $u_\rho$ (not shown) 
for $g>1$ is closely connected to polaron formation, which appears at 
about $g\simeq 1$ in the non-to-anti-adiabatic regime~\cite{FT07}. 
There the TLL typifies a (repulsive) polaronic metal~\cite{HWBAF06}. 
We emphasise that in this frequency region
the values of $K_\rho$, computed from eq.~(\ref{eqn:Krho-Sc}) via  
the static charge structure factor, reasonably agree with those 
determined by the scaling relations (\ref{eqn:krho1}), (\ref{eqn:urho1})
[see panel (a), filled and open symbols for $\omega_0=10$]. 
Furthermore, our values for the 
critical coupling, $g_c$, confirm previous 
results (although a Kosterlitz-Thouless transition is difficult 
to detect because the gap opens exponentially slow), 
as can be seen by inserting  the points where $K_\rho(g_c)=1/2$ (stars) 
into the existing phase diagram 
of the HMSF~\cite{BMH98,FHW00,HWBAF06} [cf. panel (b)]. 

Let us now look whether the situation changes when the phonon
frequency becomes smaller, i.e., when we enter the adiabatic regime. 
Figure~\ref{fig:Krho-PD}~(c) shows the scaling of $K_\rho$ at $\omega_0=0.1$, 
based on the relation~(\ref{eqn:Krho-Sc}), for up to $N=256$ sites,  
using open boundary conditions. The lines are second-order polynomial fits. 
Surprisingly, we find that $K_\rho$ scales to values smaller than unity
for any EP coupling (filled symbols). This holds for other
adiabatic phonon frequencies $\omega_0<1$ as well.   
Taking this result seriously, we arrive at the conclusion that 
the HMSF does not exhibit a metallic TLL phase with attractive interaction, 
which is in strong contradiction to the reasonings based on the 
leading-order energy scaling laws (\ref{eqn:krho1}) and 
(\ref{eqn:urho1}) [see open symbols in figure~\ref{fig:Krho-PD}~(a)].
We would like to point out, however, that our $S_c(q)$-based approach
gives apparently the correct value of the critical coupling 
for the TLL-CDW metal-insulator transition in the adiabatic HMSF. 
In previous work, $g_c$ was estimated as 
$g_c(\omega_0=0.1)\sim 2.8$~\cite{BMH98}, which is in accordance
with our DMRG-results for $K_\rho(N)$ at $g=2.81$ that clearly extrapolate 
to $K_\rho = 1/2$ in the thermodynamic limit. 
This means that the TLL-CDW transition at small $\omega_0=0.1$ 
seems to be of Kosterlitz-Thouless type as well.

To substantiate these findings, we investigate
another quantity of interest, the so-called 
momentum distribution
function,
\begin{eqnarray}
 n(k)=\frac{1}{N}\sum_{j,l}e^{ik(j-l)}
       \langle 
         c^\dagger_{j} c^{\phantom{\dagger}}_{l} 
       \rangle\,.
\end{eqnarray}
Basically $n(k)$ is the Fourier transform of the equal time Green's 
function~\cite{Gi07} and  therefore gives the occupation  
of fermionic states carrying momentum $k$. 
For free fermions, at $T=0$, all states up 
to the Fermi energy, $E_F$, are occupied, so that $n(k)$ has 
a discontinuity ($Z=1$) at the corresponding 
Fermi momentum $k_F$.\footnote{In an interacting   
Fermi liquid system there is still a discontinuity, but $Z<1$.} 
For a 1D TLL, instead of the (Fermi liquid archetypical) jump 
of $n(k)$ at $k_F$, one finds an essential power 
law singularity, corresponding to a 
vanishing quasiparticle weight $Z=0$, 
\begin{equation}
 n(k)=n_{k_{\rm F}}-C|k-k_{\rm F}|^\alpha{\rm sgn}(k-k_{\rm F})\,,
\label{eqn:singularity}
\end{equation}
where $n_{k_{\rm F}}=1/2$ for the half-filled band case.
For spinless fermions, again the critical exponent $\alpha$ is  
given by the TLL parameter $K_\rho$ :
\begin{eqnarray}\label{eqn:alpha}
 \alpha=\frac{1}{2}(K_\rho+K_\rho^{-1})-1\,.
\end{eqnarray} 
The relation~(\ref{eqn:singularity}) with~(\ref{eqn:alpha}) 
was first derived in~\cite{Lu63,LML63},
and afterwards  many analytical \cite{So79} and numerical \cite{QLSY95} 
calculations were performed in order to determine the momentum 
distribution in the weak and strong coupling regimes. 
By means of DMRG, $n(k)$ 
can be computed directly from the Fourier transformed  $\langle c^\dagger_{j} c^{\phantom{\dagger}}_{l}
        \rangle$ correlator,
$n(k) =\tfrac{1}{N}\sum_{j,l=1}^{N}\cos \left(k(j-l)\right)
        \langle 
	 c^\dagger_{j} c^{\phantom{\dagger}}_{l}
        \rangle\,,$
where $k=\frac{2\pi}{N}m$ with $m=0,\dots,N/2$. 
In the following, we calculate $n(k)$ for a linear chain with 
periodic boundary conditions, and $N=66$ sites.

\begin{figure}[tb]
\includegraphics[width=1.0\linewidth]{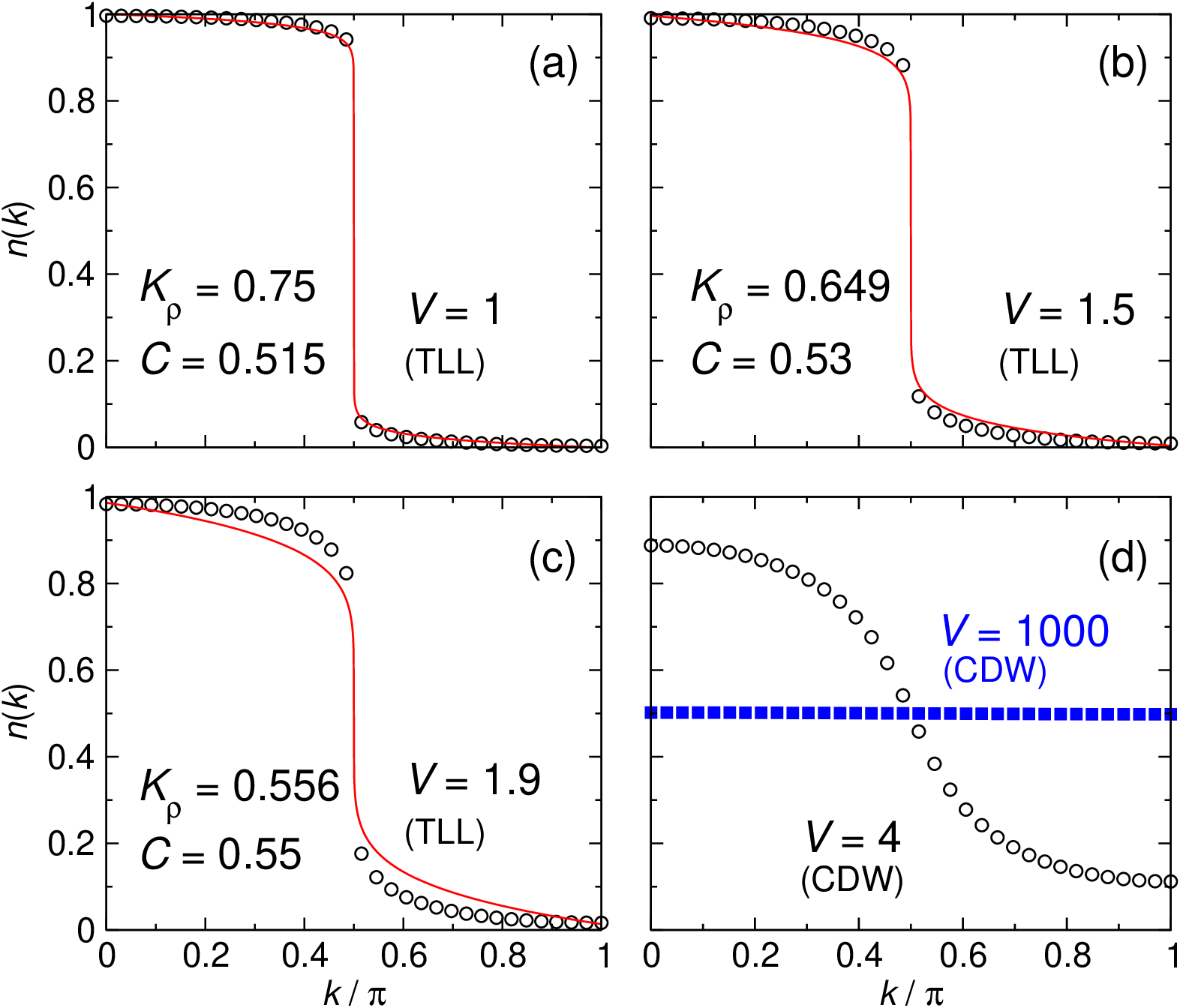}
\caption{(Colour online) Momentum distribution function $n(k)$
 for the half-filled spinless $t$-$V$ model. 
 Lines are fit to eq.~(\ref{eqn:singularity}) 
 with $K_\rho$ taken from eq.~(\ref{eqn:Krho-tV}).
}
\label{fig:nk-tV-fit}
\end{figure}
Before we discuss $n(k)$ for the HMSF, let us consider a somewhat 
simpler, purely fermionic model, however, the spinless $t$-$V$ model,  
\begin{eqnarray}
 {\cal H}
  &=& -t\sum_{j}
       \left(
	c^\dagger_{j} c^{\phantom{\dagger}}_{j+1} +
        {\rm h.c.}
       \right)
      +V\sum_j n_j n_{j+1}\,,
\label{eqn:H-tV}
\end{eqnarray} 
where $V$ is the nearest neighbour Coulomb interaction.
This is of avail because the $t$-$V$ model can also 
be mapped onto the exactly solvable XXZ-Heisenberg model
(i.e., it should exhibit the same asymptotic behaviour as the 
strong-coupling anti-adiabatic HMSF). For the $t$-$V$ model  
the analytical form of $K_\rho$ in the thermodynamic limit is known, 
\begin{eqnarray}
 K_\rho=\frac{\pi}{2{\rm arccos}[-V/(2t)]}.
  \label{eqn:Krho-tV}
\end{eqnarray}
Hence, the results obtained for $n(k)$ by DMRG can be 
fitted by the relation~(\ref{eqn:singularity}), with  
$K_\rho$ taken from eq.~(\ref{eqn:Krho-tV})~\cite{MMSSSS00}. 
Clearly, since eq.~(\ref{eqn:singularity})
is a weak-coupling result, the DMRG data for $n(k)$ 
are fitted almost perfectly for small $V$. This is demonstrated by 
figures~\ref{fig:nk-tV-fit} (a) and (b). 
Figure~\ref{fig:nk-tV-fit} (c) shows that
the agreement becomes worse for larger Coulomb interaction ($V=1.9$).
In the insulating phase ($V>2$),  the power-law singularity 
does not exist anymore 
[c.f. the smooth curves in figure~\ref{fig:nk-tV-fit} (d)].
As $V\to\infty$, the system becomes a ``perfect'' CDW, 
and consequently $n(k)=1/2$ for all momenta $k$. 
De facto this situation is realized for $V=1000$ 
already where, according to figure~\ref{fig:nk-tV-fit}~(d), 
$n(k)$ is almost constant.

\begin{figure}[tb]
\includegraphics[width=1.0\linewidth]{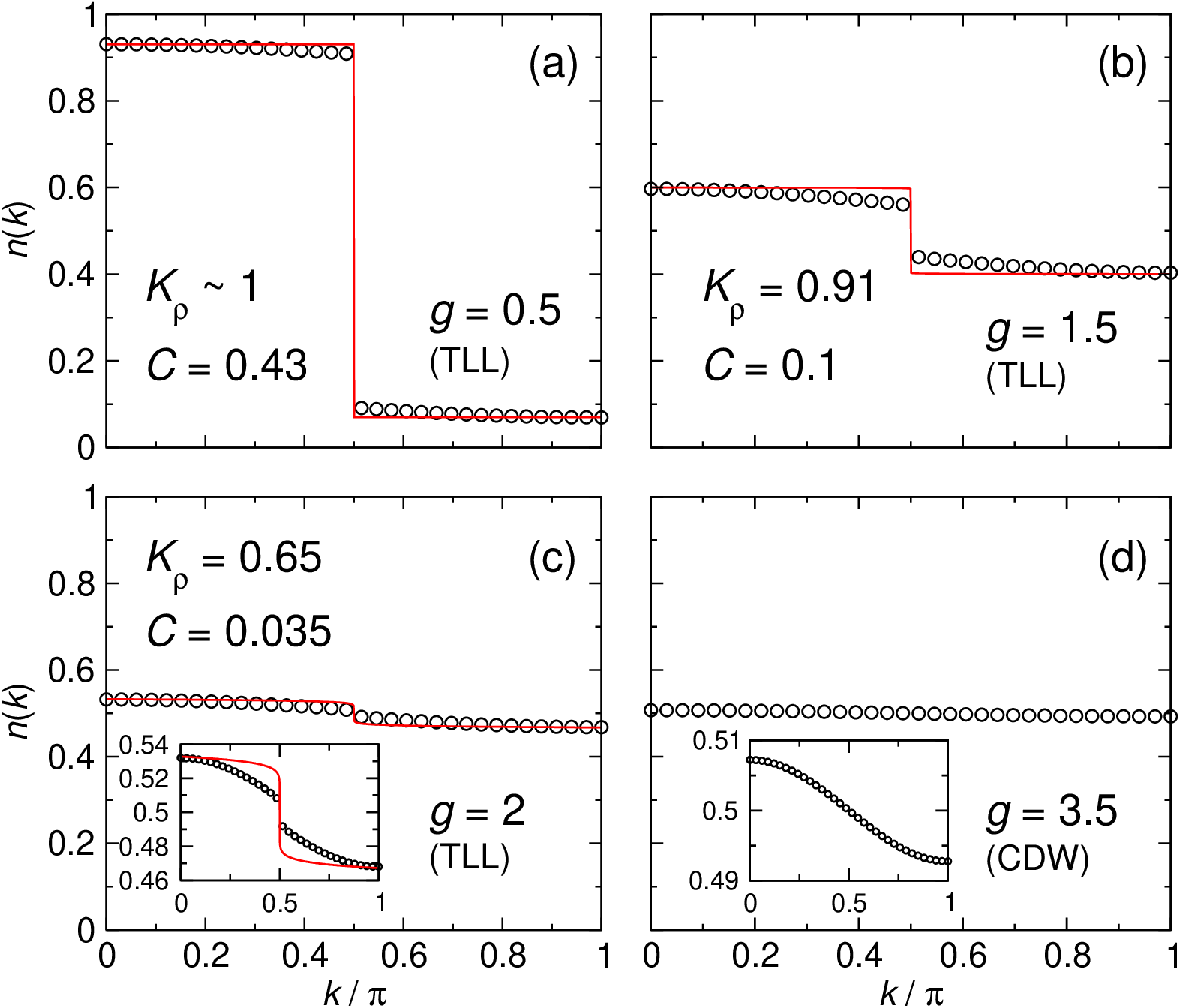}
\caption{(Colour online) Momentum distribution function $n(k)$
 in the anti-adiabatic regime ($\omega_0=10$) of the 
half-filled spinless Holstein model .
 Lines are a fit to eq.~(\ref{eqn:singularity})  
 with $K_\rho$ calculated by the DMRG.
 Insets give $n(k)$ with magnified axis of ordinate.
}
\label{fig:nk-w10}
\end{figure}

Turning now to the HMSF, we discuss at first the case
of large phonon frequencies. Figure~\ref{fig:nk-w10}
gives the $n(k)$ DMRG data obtained for $\omega_0=10$ (symbols).
Obviously, the momentum distribution is a monotonously decreasing 
function as $k$ changes from the centre ($k=0$) to the boundary 
of the Brillouin zone ($k=\pi$), with a power-law singularity at 
$k=k_{\rm F}$ in the metallic phase [panels (a) to (c)].
Quite different from the $t$-$V$ model, however, the momentum distribution 
becomes renormalised for all momenta $k$, as soon as the EP coupling is 
switched on, where $n(k)$ decreases (increases) almost uniformly
for $0\leq k < \pi/2$ ($\pi/2<k\leq\pi$). Although there is no jump 
in $n(k)$ at $k_F$, as for an ordinary Fermi liquid, for finite TLL systems
the difference $\Delta=n(k_{\rm F}-\delta)-n(k_{\rm F}+\delta)$ 
is finite (with $\delta=\pi/N=\pi/66$ in our case), and
rapidly decreases with increasing EP interaction $g$.   
As can be seen from the solid lines in figure~\ref{fig:nk-w10}
(a)-(c), the momentum distribution can be surprisingly well fitted to 
the weak-coupling result~(\ref{eqn:singularity}), just by adjusting 
the constant $C$. Thereby we take the $K_\rho$--values extracted 
from equation~(\ref{eqn:Krho-Sc}). Of course, around $k\simeq k_F$  
the agreement becomes worse as $g$ increases, but we observe 
a power-law singularity even close to the CDW transition point.
Approaching the insulating CDW state this singularity vanishes,  
and $\Delta\to 0$ as $g\to g_c$ [cf. the insets in panels (c) and (d)]. 
In the CDW phase, $n(k)\simeq 1/2$ for all $k$ [see panel (d)].
In the anti-adiabatic regime, the CDW state basically 
constitutes a polaronic superlattice, i.e the electrons are 
heavily dressed by phonons and, in addition, ordered in a A-B-structure. 
Since the polarons are self-trapped, the system tends to be a perfect CDW, 
as in the limit $V\to \infty$ of the $t$-$V$ model.

\begin{figure}[tb]
\includegraphics[width=1.0\linewidth]{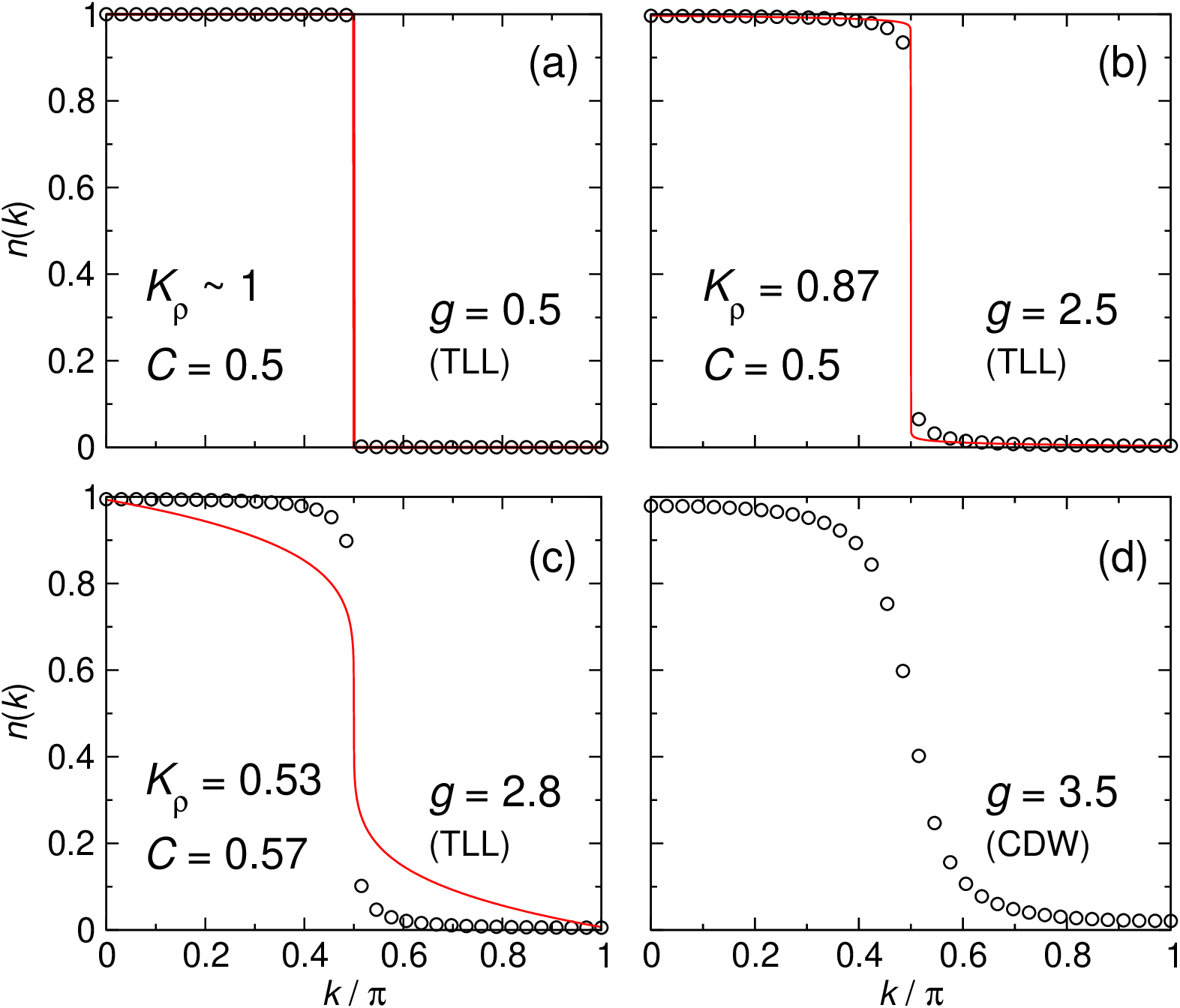}
\caption{(Colour online) Momentum distribution function $n(k)$
 in the adiabatic regime ($\omega_0=0.1$) of the 
half-filled spinless Holstein model. 
 Lines are a fit to eq.~(\ref{eqn:singularity})  
 with $K_\rho$ calculated by the DMRG.
}
\label{fig:nk-w0.1}
\end{figure}

Finally, we investigate the behaviour of $n(k)$ in the adiabatic regime  
of the HMSF (see figure~\ref{fig:nk-w0.1} for $\omega_0=0.1$). In this case, 
$n(k)$ is well approximated by equation~(\ref{eqn:singularity}) 
with $C=0.5$ for all $g<g_c$ [see panels (a) and (b)]. This means the 
weak-coupling result $K_\rho\lesssim 1$, $\Delta \lesssim 1$,
holds in (almost) the entire metallic region, where the system can be 
considered as to be composed of nearly free electrons. The momentum 
distribution starts to deviate from equation~(\ref{eqn:singularity})
just in the neighbourhood of the metal insulator transition point
$g\simeq g_c$, but even 
there $n(k)$ differs near $k\sim k_{\rm F}=\pi/2$ only. Note that $\Delta$ ($C$) 
is still very large in the transition region. Of course, very close 
to the critical point, where a strong renormalisation of
$K_\rho$ takes place (indicating the formation of a TLL with 
strong repulsive interactions), the fit of our DMRG data to the
weak-coupling relation~(\ref{eqn:singularity}) fails.
In the insulating state, $n(k)$ is given by a smooth curve 
(without power-law singularity), which -- in contrast to the 
anti-adiabatic case -- exhibits a significant curvature because the EP coupling 
used in figure~\ref{fig:nk-w0.1}~(d)  is small if compared to the half 
electronic bandwidth $2t$.\footnote{Recall 
that in the adiabatic regime $\lambda=g^2\omega_0/2t$ is the 
appropriate dimensionless EP interaction 
parameter in order to discriminate weak ($\lambda\ll 1$) and strong coupling 
($\lambda\gg 1$) situations.} Therefore system now typifies rather a  
Peierls band insulator than a polaronic superlattice.

To summarise, we investigated the properties of the metallic phase 
and the metal insulator transition in the spinless fermion Holstein model
by means of a boson pseudo-site DMRG technique supplemented by a
careful finite-size scaling analysis. In particular we determined
the Tomonaga-Luttinger correlation exponent $K_\rho$ from the long-wavelength
limit of the static charge structure factor. This approach yields 
reliable data for $K_\rho$ in the whole range of electron-phonon interaction
strengths $g$ and phonon frequencies $\omega_0$. We compare our results with 
new and previous data extracted in an alternative way from leading-order scaling  
relations for the ground-state energy and single-particle excitation
gap. In striking contrast to the latter data we find $K_\rho<1$ for all 
phonon frequencies, i.e.,  the metallic state of the HMSF represents 
a repulsive Tomonaga-Luttinger liquid, even in the adiabatic regime.
Therefore we conclude that in one dimension we have to include the 
spin degrees of freedom ~\cite{HF82,CH05,TAA05} in order to obtain 
e.g. a phase with attractive interactions  ($K_\rho>1$), or even 
dominant superconducting correlations. Furthermore, since the metal 
insulator phase boundary in the $g$-$\omega_0^{-1}$ plane obtained 
from the $K_\rho(g,\omega_0)=1/2$ line is in excellent agreement 
with previous results~\cite{BMH98,FHW00,HWBAF06}, we suppose that the 
TLL-CDW transition in the HMSF is always of Kosterlitz-Thouless type. 
Comparing the behaviour of the momentum distribution function with the 
weak-coupling TLL result reveals, however, significant differences 
regarding the nature of the metallic and insulating phases in the 
adiabatic and anti-adiabatic regimes of the HMSF. 
Whereas the metallic state is a weakly renormalised
TLL and the CDW phase typifies a Peierls band insulator at small phonon
frequencies, a polaronic metal and a polaronic superlattice are formed 
at large phonon frequencies. This is in accord with the electron and phonon
spectral properties detected in Refs.~\cite{HWBAF06,SHB06}.  
In the strong-coupling anti-adiabatic regime the momentum distribution 
function indicates perfect CDW behaviour as in the $V\to\infty$
limit of the $t$-$V$ model.
\vspace*{-0.4cm}
\acknowledgments
The authors would like to thank G.~Hager, E.~Jeckel\-mann, 
V.~Meden, and S. Nishimoto for valuable discussions. 
This work was supported by SFB 652.
\vspace*{-0.4cm}

\begin{thebibliography}{10}
\expandafter\ifx\csname url\endcsname\relax\def\url#1{\texttt{#1}}\fi

\bibitem{PN93}
\Name{Ploog K.~H. \and N\"otzel R. (Eds.)} \Book{New Concepts to fabricate
  semiconductor quantum wire and quantum dot structures} Vol. 419 of
  \emph{Lecture Notes in Physics} (Springer-Verlag, Berlin/Heidelberg) 1985.

\bibitem{IYS73}
\Name{Ishiguro T., Yamaji K. \and Saito G.} \Book{Organic Superconductors}
  (Springer-Verlag, New York) 1973;
\Name{Tsuda N., Nasu K., Yanase A. \and Siratori K.} \Book{Electronic
  Conduction in Oxides} (Springer-Verlag, Berlin) 1991;
\Name{Bishop A.~R. \and Swanson B.~I.} \REVIEW{Los Alamos Sciences}{21}{1993}{133};
\Name{Hase M., Terasaki I. \and Uchinokura K.} \REVIEW{Phys. Rev. Lett.}{70}{1993}{3651};
\Name{Wellein G., Fehske H. \and Kampf A.~P.} \REVIEW{Phys. Rev. Lett.}{81}{1998}{3956};
\Name{Toyota N., Lang M. \and M\"uller J.} \Book{Low-dimensional molecular
  metals} (Springer-Verlag, Berlin/Heidelberg) 2007.

\bibitem{Mo90}
\Name{Mott N.~F.} \Book{Metal-Insulator Transitions} (Taylor \& Francis,
  London) 1990;
\Name{Monceau P.} \REVIEW{Fiz. Tver. Tel.}{41}{1999}{759}.

\bibitem{Pe55}
\Name{Peierls R.} \Book{Quantum theory of solids} (Oxford University Press,
  Oxford) 1955;
\Name{Gr{\"u}ner G.} \Book{Density Waves in Solids} (Addison Wesley, Reading,
  MA) 1994.

\bibitem{TC03}
\Name{Takada Y. \and Chatterjee A.} \REVIEW{Phys. Rev. B}{67}{2003}{081102(R)}.

\bibitem{FWHWB04}
\Name{Fehske H., Wellein G., Hager G., Wei{\ss}e A. \and Bishop A.~R.}
  \REVIEW{Phys. Rev. B}{69}{2004}{165115};
\Name{Fehske H., Hager G. \and Jeckelmann E.} \REVIEW{Europhys. Lett.
  }{84}{2008}{57001}.

\bibitem{CH05}
\Name{Clay R.~T. \and Hardikar R.~P.} \REVIEW{Phys. Rev. Lett.
  }{95}{2005}{096401};
\Name{Hardikar R.~P. \and Clay R.~T.} \REVIEW{Phys. Rev. B}{75}{2007}{245103}.

\bibitem{TAA05}
\Name{Tezuka M., Arita R. \and Aoki H.} \REVIEW{Phys. Rev. Lett.}{95}{2005}{226401};
\Name{Tezuka M., Arita R. \and Aoki H.} \REVIEW{Phys. Rev. B}{76}{2007}{155114}.



\bibitem{Gi07}
\Name{Giamarchi T.} \Book{Quantum Physics in One Dimension} (Oxford University
  Press, Oxford) 2003.

\bibitem{Lu63}
\Name{Luttinger J.~M.} \REVIEW{J. Math. Phys.}{4}{1963}{1154}.

\bibitem{Vo95}
\Name{Voit J.} \REVIEW{Rep. Prog. Phys.}{58}{1995}{977}.

\bibitem{Ho59a}
\Name{Holstein T.} \REVIEW{Ann. Phys. (N.Y.)}{8}{1959}{325};
\Name{Holstein T.} \REVIEW{Ann. Phys. (N.Y.)}{8}{1959}{343}.

\bibitem{FHW00}
\Name{Fehske H., Holicki M. \and Wei{\ss}e A.} \REVIEW{Advances 
in Solid State Physics}{40}{2000}{235}; 
\Name{Wei{\ss}e A. \and Fehske H.} \REVIEW{Phys. Rev. B}{58}{1998}{13526}.
\bibitem{HWBAF06}
\Name{Hohenadler M., Wellein G., Bishop A.~R., Alvermann A. \and Fehske H.}
  \REVIEW{Phys. Rev. B}{73}{2006}{245120}.

\bibitem{HF82}
\Name{Hirsch J.~E. \and Fradkin E.} \REVIEW{Phys. Rev. Lett.}{49}{1982}{402};
\Name{Hirsch J.~E. \and Fradkin E.} \REVIEW{Phys. Rev. B}{27}{1983}{4302}.

\bibitem{ZFA89}
\Name{Zheng H., Feinberg D. \and Avignon M.} \REVIEW{Phys. Rev. B}{39}{1989}{9405}.

\bibitem{MHM96}
\Name{McKenzie R.~H., Hamer C.~J. \and Murray D.~W.} \REVIEW{Phys. Rev. B}{53}{1996}{9676}.

\bibitem{BMH98}
\Name{Bursill R.~J., McKenzie R.~H. \and Hamer C.~J.} \REVIEW{Phys. Rev. Lett.}{80}{1998}{5607}.


\bibitem{SHBWF05}
\Name{Sykora S., H{\"u}bsch A., Becker K.~W., Wellein G. \and Fehske H.}
  \REVIEW{Phys. Rev. B}{71}{2005}{045112}.

\bibitem{KT73}
\Name{Kosterlitz J.~M. \and Thouless D.~J.} \REVIEW{J. Phys. C}{6}{1973}{1181}.

\bibitem{Ha80}
\Name{Haldane F. D.~M.} \REVIEW{Phys. Rev. Lett.}{45}{1980}{1358}.

\bibitem{To50}
\Name{Tomonaga S.} \REVIEW{Prog. Theor. Phys.}{5}{1950}{544}.

\bibitem{Ca84}
\Name{Cardy J.~L.} \REVIEW{J. Phys. A}{17}{1984}{L385}.

\bibitem{FK90}
\Name{Frahm H. and Korepin V. E.} \REVIEW{Phys. Rev. B}{42}{1990}{10553};
\Name{Kawakami N. and Yang S. K.} \REVIEW{Phys. Lett. A}{148}{1990}{359}.

\bibitem{Og91}
\Name{Ogata M., Luchini M. U., Sorella S., and Assaad F. F.} 
\REVIEW{Phys. Rev. Lett.}{66}{1991}{2388}.

\bibitem{WFWB00}
\Name{Wei{\ss}e A., Fehske H., Wellein G. \and Bishop A.~R.} \REVIEW{Phys. Rev.
  B}{62}{2000}{R747};
\Name{Fehske H., Wellein G., Hager G., Wei{\ss}e A., Becker K.~W. \and Bishop
  A.~R.} \REVIEW{Physica B}{359--361}{2005}{699}.

\bibitem{Fi75}
\Name{Firsov Y.~A.} \Book{Polarons} (Izd. Nauka, Moscow) 1975.

\bibitem{FT07}
\Name{Fehske H. \and Trugman S.~A.} \Book{Numerical solution of the Holstein
  polaron problem} in \Book{Polarons in Advanced Materials}, edited by
  \Name{Alexandrov A.~S.} Vol. 103 of \emph{Springer Series in Material
  Sciences} (Canopus/Springer Publishing, Dordrecht) 2007 pp. 393--461.

\bibitem{SHB06}
\Name{Sykora S., H{\"u}bsch A. \and Becker K.~W.} \REVIEW{Europhys. Lett.}{76}{2006}{644}.

\bibitem{EGN05}
\Name{Ejima S., Gebhard F. \and Nishimoto S.} \REVIEW{Europhys. Lett.}{70}{2005}{492}.

\bibitem{EHF09}
\Name{Ejima S., Hager G. \and Fehske H.} \REVIEW{Phys. Rev. Lett.}{102}{2009}{106404}.

\bibitem{CSC99}
\Name{Clay R.~T., Sandvik A.~W. \and Campell D.~K.} \REVIEW{Phys. Rev. B}{59}{1999}{4665}.

\bibitem{JW98b}
\Name{White S.~R.} \REVIEW{Phys. Rev. Lett.}{69}{1992}{2863};
\Name{Jeckelmann E. \and White S.~R.} \REVIEW{Phys. Rev. B}{57}{1998}{6376}; 
\Name{Jeckelmann E. \and Fehske H.} \REVIEW{Rivista del Nuovo Cimento}{30}{2007}{259}.


\bibitem{LML63}
\Name{Mattis D. C. and Lieb E. H.} 
\REVIEW{J. Math. Phys.}{6}{1965}{304}.


\bibitem{So79}
\Name{S\'{o}lyom J.} \REVIEW{Adv. Phys.}{28}{1979}{201};
\Name{Ogata M. \and Shiba H.} \REVIEW{Phys. Rev. B}{41}{1990}{2326};
\Name{Pruschke T. \and Shiba H.} \REVIEW{Phys. Rev. B}{44}{1991}{205}.

\bibitem{QLSY95}
\Name{Qin S., Liang S., Su Z. \and Yu L.} \REVIEW{Phys. Rev. B}{52}{1995}{5475}.
\bibitem{MMSSSS00}
\Name{Meden V., Metzner M., Schollw\"ock U., Schneider O., Stauber T.,
and Sch\"onhammer K.} \REVIEW{Eur. Phys. J. B}{16}{2000}{631}.
\end{thebibliography}

\end{document}